\documentclass[10pt]{article}
\usepackage[utf8]{inputenc}
\usepackage[T1]{fontenc}
\usepackage{amsmath}
\usepackage{amsfonts}
\usepackage{amssymb}
\usepackage[version=4]{mhchem}
\usepackage{stmaryrd}
\usepackage{enumitem}
\usepackage{setspace}
\usepackage[hyphens]{url}
\usepackage{hyperref}
\usepackage[a4paper,margin=3cm]{geometry}
\hypersetup{breaklinks=true}
\usepackage{graphicx}
\usepackage{pdflscape}
\begin{document}

\noindent
\begin{center}
    \LARGE \textbf{A Pigouvian Matchmaker Mechanism for De-escalating the AGI Race}
\end{center}

\noindent\makebox[\linewidth]{\rule{\textwidth}{1pt}}

\begin{center}
    \large Eduard Kapelko \\
\end{center}

\vspace{1em}

\section*{Abstract}
A formal mechanism is presented in which a willing regulator-matchmaker fosters cooperation on resources among participants in the AGI race, collects a Pigouvian tax based on the speed-up it induces, and invests the proceeds into alignment research. The construction is derived in the continuous-time options framework of Tan (2025) in which cooperation is treated as a jump in the underlying asset value of participating players, the Pigouvian component is matched to the marginal effect of increasing expected loss, and the total collected fund endogenizes the rate of learning on safety. It is shown how the framework allows for determining participation and optimal activity levels.

Conditions under which it is optimal to enter the market are derived, and it is proven that if the orthogonality condition holds between the supported portfolio and the abilities component, the Suicide Region collapses at finite time, and the upper bound for this time is derived as sum of a deterministic and random term. Finally, if orthogonality is violated, it is proven that enhancing matchmaker capacity does not recover the market's superiority. The construction links research areas including two-sided markets, Pigouvian taxes, self-regulatory organizations, private law enforcement, evolutionary modeling of AI races, real options and option games, measurement of comparative progress and analysis of the Suicide Region.

\section*{Motivation}
Tan (2025) sets up the AGI race as a symmetric preemption game with continuous time, developing the earlier framework of Armstrong, Bostrom \& Shulman (2016). Each player's asset value is described by GBM and the safety learning function is determined by the length of time safety research takes before the tool is released and by an exogenous learning rate.

The leader and the follower have identical payoff functions which include an expectation loss component, stemming from a misalignment. Given the condition for equal payoffs the term drops out, and the preemption threshold turns out to be independent of this expectation loss. However, the survival threshold contains the expectation loss, and at large losses the preemption threshold falls below the survival threshold. The Suicide Region is defined as the interval of asset values within which the game prescribes a race even when the risk-adjusted present value is negative.

There are identified three types of interventions:
\begin{itemize}
    \item The privatization of liability (increase in preemption threshold)
    \item Windfall clauses (O'Keefe et al. 2020) (equal split of the gains for follower and leader)
    \item Sprint verification
\end{itemize}

All these require an exogenous agent with coercive abilities, which in turn runs into the problem that no global sovereign exist. Similar approaches (e.g., Han et al. 2020, 2021; Cimpeanu et al. 2022) similarly rely on a regulator, but cannot answer the most important question of: how can we afford and legalize a regulator when no central authority exists?

We propose a new mechanism that does not involve coercion (participation is voluntary), is self-financed through the race surplus itself, the regulator is not involved in the race (this is ensured by transparency and a democratic governance system), and funds accumulated allow endogenous increases of the learning rate, hence downward shift of the survival threshold, and eventual closure of the Suicide Region.

\section*{Formal Model}
Let $N \ge 2$ be the number of participants in the AGI race, indexed $i \in \{1, \dots, N\}$. Each player is associated with a process $V_t^i$:
$$dV_t^i = \mu V_t^i dt + \sigma V_t^i dZ_t^i, \qquad \mu = r_f - \delta.$$

The safety learning function is given by $\pi(\tau) = 1 - e^{-\lambda \tau}$, where $\tau$ is the duration of safety research prior to deployment, and $\lambda > 0$ denotes the exogenous rate. All players have the same sunk cost $I$ and face a common $D$.

Leader payoff:
$$L(V_\tau, \tau) = (1-S)\pi(\tau)V_\tau - (1-\pi(\tau))D - I,$$
Follower payoff:
$$F(V_\tau, \tau) = S\pi(\tau)V_\tau - (1-\pi(\tau))D.$$

Preemption threshold:
$$V_P^* = \frac{I}{(1-2S)\pi(\tau)},$$

Survival threshold:
$$V_S^* = \frac{I + (1-\pi(\tau))D}{(1-S)\pi(\tau)}.$$

Time until deployment of player $i$ is denoted $\tau_i$, while deployment is triggered when $V_t^i$ crosses the strategically active threshold (in Tan's equilibrium this is $V_P^-$, in social optimum $V_S^-$).

Assume that any pair $(i, j)$ may at time $t$ enter into a cooperative deal. We observe a simultaneous jump in both asset values:
$$V_t^i \to V_t^i + \tfrac{1}{2}(k-1)(V_t^i + V_t^j), \qquad V_t^j \to V_t^j + \tfrac{1}{2}(k-1)(V_t^i + V_t^j),$$

where $k > 1$ is the synergy multiplier. To simplify modeling, we assume the total increment $(k-1)(V_t^i + V_t^j)$ is distributed symmetrically and includes joint use of compute, exchange of data, and exchange of non-conflicting architectural discoveries, raising the capitalized value of each participant without splitting it into shares.

Direct cooperation without an intermediary requires payment of search costs $s > 0$. In context, search means not so much finding a partner and establishing contact (which is non-critical for a small number of participants), but rather finding compatible sets of complementary assets (datasets, compute, non-conflicting architectural discoveries) within each lab's declared portfolio. Also involved are agreement on the structure of IP sharing and the gains from joint work, legal formalization of the deal without disclosing competitively-sensitive information prior to signing, and the establishment of bilateral verification infrastructure for compliance, not relying on a third-party arbiter. Assume that aggregate $s$ is sufficiently large: at typical $(V_t^i, V_t^j)$
$$s > \beta \cdot (k-1)(V_t^i + V_t^j), \quad \beta \in (0,1),$$

that is, direct cooperation is unprofitable at the current equilibrium and empirically rare (Williamson 1979 on transaction costs; Enabling Frontier Lab Coordination 2025 on the current state of inter-lab interaction). This is consistent with small $N$ as an enforcement mechanism: a small number of labs ensures identification of potential partners and observability of agreement violations, but does not eliminate the costs of establishing and verifying each specific deal.

Cooperation is socially negative without internalization of the externality. Increase of $V_t^i, V_t^j$ brings the crossing of the fixed thresholds $V_P^-, V_S^-$ closer, but does not shift the thresholds themselves. Consequently, expected $\tau$ until deployment shrinks, $\pi(\tau)$ decreases, and $(1-\pi(\tau))D$ grows. If this acceleration is not internalized via a fee, cooperation increases aggregate expected damage.

The condition for the non-emptiness of the Suicide Region $\mathcal{S} = \{V_P^- < V < V_S^-\}$ is derived directly from $V_P^- < V_S^-$:
\begin{equation*}
\frac{I}{(1-2S)\pi} < \frac{I + (1-\pi)D}{(1-S)\pi} \quad \Longleftrightarrow \quad D > \frac{IS}{(1-2S)(1-\pi)}. \tag{$\ast$}
\end{equation*}

The parametric subset
$$\mathcal{S}(\pi, I, D, S) = \{V : V_P^-(\pi) < V < V_S^-(\pi)\}$$

is non-empty if and only if condition $(\ast)$ holds.

In the ``winner-takes-all'' limit ($S \to 0$), condition $(\ast)$ reduces to $(1-\pi)D > 0$, i.e., the Suicide Region is non-empty for any non-zero expected damage and any $\pi < 1$.

\section*{Proposed Mechanism}
The regulator-matchmaker $M$ pairs labs for resource cooperation such as joint compute use, data exchange, and exchange of non-conflicting architectural discoveries, after which it levies a fee $\tau_M = \delta_M + r$ per deal. The part $\delta_M$ covers operational cost and is assumed negligibly small hereafter for simplicity; the part $r$, the Pigouvian component, is directed into alignment research. The accumulated fund $R(t)$ raises the safety learning rate $\lambda$, which shifts the survival threshold $V_S^-$ downward. At sufficient intensity, $V_S^-$ descends to $V_P^-$ and the Suicide Region closes.

For such a construction to work, three conditions must hold.
\begin{itemize}
    \item First, it must be profitable for participants to go through $M$ rather than forgoing cooperation or seeking a partner around it.
    \item Second, the fee $r$ must be calibrated against the race acceleration that cooperation introduces, otherwise the mechanism subsidizes the very damage it works against.
    \item Third, the accumulated fund must convert into $\lambda$ rather than leaking into capability-drift $\mu$.
\end{itemize}
This section develops the first two conditions and formalizes the dynamics of $\lambda(t)$ under the assumption of benign conversion; the third condition is left for \S7.

Consider player $i$'s decision at time $t$ given an available pair $j$. Cooperation generates a jump of total value $(k-1)(V_t^i + V_t^j)$, distributed symmetrically among participants: labs jointly use compute, exchange datasets, share architectural results that are not in direct competitive conflict. The player has three options, each with its own net jump $\Delta V_i$:

$$\Delta V_i^{coop,M} = \tfrac{1}{2}\bigl[(k-1)(V_t^i + V_t^j) - (\delta_M + r)\bigr],$$

$$\Delta V_i^{coop,direct} = \tfrac{1}{2}\bigl[(k-1)(V_t^i + V_t^j) - s\bigr],$$

$$\Delta V_i^{none} = 0.$$

Direct cooperation without an intermediary requires search costs $s$, empirically high --- in the sense developed in \S2: not absence of knowledge about partners' existence, but the laboriousness of finding specific complementarities, agreeing IP-sharing, legal formalization, and bilateral verification without a third-party arbiter. Williamson (1979) provides the theoretical framework for these transaction costs; Enabling Frontier Lab Coordination (2025) provides a recent empirical estimate of their magnitude for frontier labs. The regulator reduces them to $s_M \ll s$, in the limit to zero, since it specializes precisely in these four subtasks. Then participation in the $M$-deal dominates direct cooperation when
\begin{equation*}
\tau_M = \delta_M + r < s \tag{P1}
\end{equation*}
and dominates declining cooperation when
\begin{equation*}
(k-1)(V_t^i + V_t^j) > \delta_M + r. \tag{P2}
\end{equation*}

At typical $V_t^i, V_t^j$ of large labs and moderate values of $\tau_M$, both conditions hold with substantial margin. This is the basic operational hypothesis of the work: for participants with large capitalization, the cooperation gain substantially exceeds any reasonable fee, and the remaining substantive question is the calibration of $r$.

Cooperation is socially negative without internalization of the externality. An upward jump in $V_t^i, V_t^j$ brings the crossing of the fixed thresholds $V_P^-, V_S^-$ closer without shifting the thresholds themselves: expected time until deployment shrinks, $\pi(\tau)$ decreases, expected damage $(1-\pi)D$ grows. This is the externality to be internalized. Quantitatively it can be estimated as follows: for a process $V_t$ following geometric Brownian motion with $\mu > \sigma^2/2$, the expected first-passage time to a threshold decreases in the current value with first-order elasticity $\partial \mathbb{E}\tau / \partial V \approx -1/(\mu V_t)$, to first order in $\sigma$. The total jump $\Delta V_{coop} = (k-1)(V_t^i + V_t^j)$ in a deal between $i$ and $j$ shortens the expected time until deployment by $\Delta\tau \approx -2(k-1)/\mu$, and the corresponding marginal increment in expected damage, by the chain rule, equals

$$\Delta\mathrm{SC} = -\frac{\partial \pi}{\partial \tau} \cdot \Delta\tau \cdot D = \lambda(t) e^{-\lambda(t)\tau} \cdot \frac{2(k-1)}{\mu} \cdot D.$$

By the Pigouvian taxation principle (Pigou 1920; Sandmo 1975), the fee must equal the marginal social damage:
\begin{equation*}
r^* = \lambda(t) e^{-\lambda(t)\tau} \cdot D \cdot \frac{2(k-1)}{\mu}. \tag{P3}
\end{equation*}

In a simplified intuitive form, (P3) reads as $r^* \approx (\partial p_r / \partial C) \cdot D \cdot \Delta C$: the fee is proportional to the marginal increase in the probability of misalignment per unit increase in capabilities. In Tan's formalism this increase is expressed through the derivative of the safety function with respect to time and through the shortening of waiting time induced by the GBM jump. Aggregately, the expected increment $(1-\pi)D$ caused by the deal is fully covered by the levied fee.

The fees aggregate into a fund

$$R(t) = r \int_{t_0}^t M(s) ds,$$

where $M(s)$ is the flow of match-deals. In Tan, the safety learning rate $\lambda$ is exogenous; in our setting, the fund finances alignment research, and we endogenize $\lambda$ via
\begin{equation*}
\lambda(t) = \lambda_0 + \alpha R(t), \qquad \dot\lambda(t) = \alpha r M(t), \tag{3.1, 3.2}
\end{equation*}
where $\alpha > 0$ is the efficiency of converting alignment capital into safety learning rate. Under a sustained flow $M(t) \to \bar M$ this yields linear growth $\lambda(t) \sim \lambda_0 + \alpha r \bar M (t - t_0)$ and, consequently, $\pi(\tau; t) \to 1$ as $t \to \infty$ for any fixed $\tau > 0$.

Equation (3.1) requires two caveats. The first concerns the interpretation of $\pi$. Canonically $\pi(\tau) = 1 - e^{-\lambda\tau}$ is the probability of safe deployment given fixed duration $\tau$ and constant $\lambda$. With time-varying $\lambda(t)$, two interpretations are possible.
\begin{itemize}
    \item The myopic one applies the current $\lambda(t)$ retroactively to a fixed $\tau$, $\pi(\tau; t) = 1 - e^{-\lambda(t)\tau}$, and ignores the fact that past smaller values of $\lambda(s)$ for $s < t$ contributed less to accumulated safety.
    \item The integral one uses the hazard-rate representation $\pi(t) = 1 - \exp(-\int_{t_0}^t \lambda(s) ds)$, internally consistent and recovering the myopic one under constant $\lambda$.
\end{itemize}

The myopic overstates $\pi$ under growing $\lambda$, and hence myopic estimates of Suicide Region closure times are lower bounds relative to the integral ones. The main text below is written in myopic notation for compatibility with the formalism; central claims are additionally verified in the integral form.

The second caveat concerns the scalar $\alpha$ itself. Contemporary alignment research suffers from a dual-use problem: part of the funded work, nominally directed into $\lambda$, actually shifts $\mu$. This means $\alpha$ in (3.1) implicitly assumes a portfolio orthogonal to the capabilities component in a sense we make precise later through the correlation metric $\rho$ of Ren et al. (2024). Without orthogonality, equation (3.1) loses its sign. In general, orthogonality is taken as an idealization as gives it operational content and delineates the regime in which it is applicable.

With these caveats, conditions (P1), (P2), (P3) jointly characterize the mechanism at the micro level: they specify when a participant goes through $M$ and how much $M$ collects from them. The next section verifies that these conditions survive the transition to the equilibrium of the repeated game, that is, that participants do not learn to bypass the regulator after the first match.

\section*{Mechanism Robustness in Equilibrium}
We showed that in a single round, participation in the $M$-deal is dominant under (P1--P3). But participants do not play a single round. A pair $(i, j)$, having once established contact through the regulator, can in the future cooperate directly, saving $\tau_M$ net of residual search costs. If such bypasses become equilibrium, the fund $R(t)$ does not fill, and the central scheme collapses.

This is a standard private-enforcement problem, and it has a canonical answer. In a discounted repeated game with discount factor $\gamma \in (0,1)$ and expected flow $\mu_M$ of match-deals per unit time, the per-period payoff of a participant in terms of $V_t$-jumps is

$$\Pi_{period} \approx \mu_M \bigl[(k-1)\bar V - \tfrac{1}{2}\tau_M\bigr],$$

where $\bar V = (V_t^i + V_t^j)/2$. The one-shot gain from bypassing in a single deal is the saved half of the fee, $\tau_M/2$. The cost of bypassing, if the regulator excludes violators upon detection, is the present value of the foregone deal flow,

$$\Phi(\gamma, \mu_M) = \mu_M \bigl[(k-1)\bar V - \tfrac{1}{2}\tau_M\bigr] \cdot \frac{\gamma}{1-\gamma}.$$

The condition for suppressing bypass:
\begin{equation*}
\frac{\tau_M}{2} < \mu_M \bigl[(k-1)\bar V - \tfrac{1}{2}\tau_M\bigr] \cdot \frac{\gamma}{1-\gamma}. \tag{P4}
\end{equation*}

At typical $\gamma \to 1$, $\mu_M \ge 1$, $(k-1)\bar V \gg \tau_M$, this condition holds with substantial margin. Here we have a formal justification of the intuition about small $N$ as an enforcement mechanism: a small number of frontier labs implies a high effective $\gamma$, because future deals within a small group are nearly guaranteed (Bernstein 1992 on the Diamond District; Greif 1993 on the Maghribi). Transparency of $M$'s decisions and financial flows, developed in \S6, ensures observability of violations: bypass becomes common knowledge, and the exclusion threat is backed by coordination in excluding behavior of the remaining participants.

Here arises a second question: the standard two-sided platforms literature (Caillaud \& Jullien 2003; Rochet \& Tirole 2003) indicates that below critical mass such platforms die: network ties do not form, network effects do not activate, platform value collapses to zero. This creates a classic bootstrap problem: even a socially useful platform may fail to launch if the first participants lack the incentive to join.

The proposed mechanism differs qualitatively. The Pigouvian component $r$ is directed into alignment per transaction, not proportionally to total volume or network effect. Each individual match-deal, including the first, strictly increases $\lambda(t)$:

$$\frac{\partial \lambda(t)}{\partial M(t)} = \alpha r > 0 \quad \forall M(t) \ge 1.$$

Hence the regulator brings social benefit from the moment of formation: a single match-deal suffices for $\dot\lambda > 0$, which begins shifting $V_S^-$ downward. Critical mass determines the completeness of the deterrence effect --- how firmly bypass is blocked (P4) --- not the presence of a safety effect as such. Below critical mass, the exclusion threat is weak, bypass is possible, but even a marginally successful regulator remains net positive in terms of social damage.

This argument, however, requires care. All three consequences --- bootstrap from one match-deal, monotonic growth of $\lambda$, net-positive contribution of each transaction --- rest on the positive sign of $\partial\lambda/\partial M$. Under dual-use this sign is not guaranteed: part of the fund, nominally directed into alignment, actually shifts not $\lambda$ but the $\mu$ of the process $V_t$. When such leakage is large, each additional match-deal accelerates capability more than it raises safety, and the regulator works in the minus. The bootstrap argument holds when the funded portfolio is orthogonal to the capabilities component, and does not hold outside this regime. That is, the early-stage binding constraint is not ``finding sufficient seed capital'' but ``finding a seed portfolio that does not accelerate the race,'' and calibration of the matchmaker's throughput without control over this condition is useless.

Classic solutions to the critical-mass problem --- subsidizing early participants (CAIF, LTFF, Open Philanthropy), cross-side network effects (compute access, liability shield), coordination signals (Frontier Model Forum, MLCommons) --- remain applicable as accelerators but not as necessary launch conditions. The very fact of the first deal gives a positive contribution to $\lambda$ if that deal finances an orthogonal portfolio; without orthogonality the same network effect works on race acceleration. The next section proves that within the orthogonal regime the Suicide Region closes in finite time; the section after formalizes what makes the regime orthogonal and what happens upon exiting it.

\section*{Closing the Suicide Region}
The mechanism passes muster only if it actually closes the Suicide Region in finite time, otherwise the entire construction remains a rhetorical gesture. This section proves the corresponding statement in three layers: a deterministic estimate of the closure time, a stochastic estimate of the probability of premature entry of $V_t$ into the region before its closure, and a formulation of the condition under which the fund attains the critical safety learning rate earlier than the capability leader crosses the preemption threshold.

The canonical safety function $\pi(\tau) = 1 - e^{-\lambda\tau}$ specifies the probability of safe deployment under fixed duration $\tau$ and constant $\lambda$. With endogenously varying $\lambda(t)$, the correct form is the hazard-rate integral $\pi(t) = 1 - \exp(-\Lambda(t))$, $\Lambda(t) = \int_{t_0}^t \lambda(s) ds$, recovering the canonical expression under constant $\lambda$. The myopic form $\pi(\tau;t) = 1 - e^{-\lambda(t)\tau}$, applying the current $\lambda(t)$ retroactively, overstates $\pi$ under growing $\lambda$ and gives lower bounds on closure times.

We fix the assumptions:
\begin{itemize}
    \item (A1) The function $\lambda: [t_0, \infty) \to \mathbb{R}_+$ is continuous, $\lambda(t_0) = \lambda_0 \ge 0$.
    \item (A2) The flow of match-deals $M(t) \ge \bar M > 0$ for all $t \ge t_0$.
    \item (A3) Parameters $\alpha, r > 0$ are fixed.
    \item (A4) The fund flow $rM(t)$ bifurcates into safety- and capability- components as
    \begin{equation*}
    \dot\lambda(t) = \alpha(1-\rho)rM(t), \qquad \mu(t) = \mu_0 + \kappa\rho rM(t), \tag{5.0}
    \end{equation*}
    where $\rho \in [0,1]$ is the correlation of the funded work with the first principal component of the ``model $\times$ benchmark'' space (Ren et al. 2024), $\kappa \ge 0$ is the leakage gain into drift, and $\rho < \rho_{crit}$ (explicit form --- formula 6.3 below). At $\rho = 0$ the pure dynamics of \S3 is recovered.
    \item (A5) $S \in (0, 1/2)$, asymmetric ``leader--follower'' structure.
\end{itemize}

Under (A1)--(A5), from (5.0) we obtain the lower bound
\begin{equation*}
\dot\Lambda(t) = \lambda(t) \ge \lambda_0 + \alpha(1-\rho)r\bar M(t - t_0). \tag{5.1}
\end{equation*}

At each $t$ with $\pi(t) \in (0,1)$, time-dependent thresholds are defined
$$V_P^-(t) = \frac{I}{(1 - 2S)\pi(t)}, \qquad V_S^-(t) = \frac{I + (1 - \pi(t))D}{(1 - S)\pi(t)},$$

and the Suicide Region $\mathcal{S}(t) = (V_P^-(t), V_S^-(t))$, if non-empty. The closure of $\mathcal{S}$ is conveniently rewritten as a condition on $\pi$ directly.

\textit{Lemma 1 (algebraic equivalence).} $\mathcal{S}(t) \ne \emptyset$ if and only if $(1 - \pi(t))D > IS/(1 - 2S)$.

\textit{Proof.} $V_P^-(t) < V_S^-(t)$ is equivalent to
$$\frac{I}{(1-2S)\pi} < \frac{I + (1-\pi)D}{(1-S)\pi}.$$

Multiplying by $\pi(1-S)(1-2S) > 0$ we obtain $I(1-S) < (1-2S)I + (1-2S)(1-\pi)D$, whence $IS < (1-2S)(1-\pi)D$. $\blacksquare$

\textit{Lemma 2 (monotonicity of $\pi$).} Under (A1)--(A4) the function $t \mapsto \pi(t)$ is strictly increasing on $[t_0, \infty)$ and $\pi(t) \to 1$ as $t \to \infty$.

\textit{Proof.} $\dot\pi(t) = \lambda(t)e^{-\Lambda(t)} > 0$ under (A2)--(A4). From (5.1) it follows that $\Lambda(t) \ge \lambda_0(t-t_0) + \tfrac{1}{2}\alpha(1-\rho)r\bar M(t-t_0)^2 \to \infty$, whence $\pi(t) \to 1$. $\blacksquare$

Combining the two lemmas yields the main statement.

\textit{Theorem 1 (closure of the Suicide Region).} Under (A1)--(A5) there exists $t^* < \infty$ such that $\mathcal{S}(t) = \emptyset$ for all $t \ge t^*$. In integral form
\begin{equation*}
t^* \le t_0 + \frac{-\lambda_0 + \sqrt{\lambda_0^2 + 2\alpha(1-\rho)r\bar M L^*}}{\alpha(1-\rho)r\bar M}, \qquad L^* = \left[\ln\frac{D(1-2S)}{IS}\right]_+; \tag{5.2}
\end{equation*}
in myopic form
\begin{equation*}
t^* \le t_0 + \frac{1}{\alpha(1-\rho)r\bar M}\left[\frac{L^*}{\tau} - \lambda_0\right]_+. \tag{5.3}
\end{equation*}

\textit{Proof.} When $D(1-2S) \le IS$ the case is trivial: $L^* = 0$, the condition of Lemma 1 fails already at $\pi = 0$, and $\mathcal{S}(t) = \emptyset$ for all $t$. In the non-trivial case $L^* > 0$, closure by Lemma 1 is equivalent to $\Lambda(t) \ge L^*$. From (5.1) we obtain the quadratic inequality
$$\lambda_0(t-t_0) + \tfrac{1}{2}\alpha(1-\rho)r\bar M(t-t_0)^2 \ge L^*;$$
the positive root gives (5.2). The myopic estimate follows from the requirement $\lambda(t)\tau \ge L^*$ with linear lower bound $\lambda(t) \ge \lambda_0 + \alpha(1-\rho)r\bar M(t-t_0)$. $\blacksquare$

The behavior of estimate (5.2) unfolds in two regimes. When the seed rate $\lambda_0$ is large relative to the mechanism's contribution, $\lambda_0^2 \gg 2\alpha(1-\rho)r\bar M L^*$, the expression under the root is dominated by the first term, and $t^* - t_0 \approx L^*/\lambda_0$ --- a linear dependence on the inverse seed rate. When the seed rate is small, $\lambda_0^2 \ll 2\alpha(1-\rho)r\bar M L^*$, the second term dominates, and $t^* - t_0 \approx \sqrt{2L^*/[\alpha(1-\rho)r\bar M]}$ --- a square-root dependence on throughput. The second regime (the one the mechanism design is intended for, $\lambda_0$ small, $\bar M$ substantial), and in it doubling the matchmaker's throughput shortens $t^*$ only by a factor of $\sqrt{2} \approx 1.41$, not by half.

We find it necessary to make a remark on the boundary of applicability of these results. The thresholds $V_P^-(t), V_S^-(t)$ are derived from instantaneous comparison of payoffs at the moment of deployment under current $\lambda(t)$, without accounting for the option value of waiting under growing $\lambda$. In the full option game with time-varying $\lambda$, the thresholds $\tilde V_P^-(t), \tilde V_S^-(t)$ are solutions to an HJB problem with non-constant coefficients under strategic coordination of leader and follower; monotonicity of the waiting option predicts $\tilde V_P^- \ge V_P^-$, and standard convexity of the value function (Grenadier 2002) gives closure of $\tilde{\mathcal{S}}$ no later than the static $\mathcal{S}$ under constant parameters. Technically, the transfer to non-stationary $\lambda(t)$ in a strategic environment requires a separate proof, since under strong asymmetry $\tilde V_P^- < V_P^-$ is in principle not excluded.

Closure of $\mathcal{S}$ as a set in $V$-space is necessary but not sufficient. A situation is possible in which $V_t$ falls into $\mathcal{S}(t)$ at some $t < t^*$, and the racing equilibrium triggers deployment before the mechanism closes the region. The stochastic complement estimates the probability of such entry. Denote $\rho_{\mathcal{S}}(t_0; T) = \mathbb{P}[\exists t \in [t_0, T] : V_t \in \mathcal{S}(t)]$.

\textit{Lemma 3 (boundary estimate).} $\rho_{\mathcal{S}}(t_0; t^*) \le \mathbb{P}[\tau_{V_P^-(t^*)} \le t^*]$, where $\tau_a = \inf\{t \ge t_0 : V_t \ge a\}$.

\textit{Proof.} By Lemma 2, $\pi(t)$ is increasing, whence $V_P^-(t) = I/[(1-2S)\pi(t)]$ is decreasing. The minimum of $V_P^-(t)$ on $[t_0, t^*]$ is attained at $t = t^*$. Entry into $\mathcal{S}(t)$ requires $V_t \ge V_P^-(t) \ge V_P^-(t^*)$, so the event is a sub-event of $\{\tau_{V_P^-(t^*)} \le t^*\}$. $\blacksquare$

The standard first-passage formula for GBM (Karatzas \& Shreve 1991, \S3.5.C; Borodin \& Salminen 2002, II.1.1.4) with $V_{t_0} = v < a$ gives
\begin{equation*}
\mathbb{P}[\tau_a \le T] = \Phi\left(\frac{\ln(v/a) + \mu' T}{\sigma\sqrt{T}}\right) + \left(\frac{v}{a}\right)^{1 - 2\mu/\sigma^2} \Phi\left(\frac{\ln(v/a) - \mu' T}{\sigma\sqrt{T}}\right), \tag{5.4}
\end{equation*}
where $\mu' = \mu - \sigma^2/2$, and the exponent in the second term is written via $\mu$ without prime ($-2\mu'/\sigma^2 = 1 - 2\mu/\sigma^2$ --- the canonical Karatzas--Shreve form). Substituting into Lemma 3:

\textit{Theorem 2 (stochastic complement).} Under (A1)--(A5) and provided $V_{t_0} < V_P^-(t^*)$
\begin{equation*}
\rho_{\mathcal{S}}(t_0; t^*) \le \Phi(d_+) + \left(\frac{V_{t_0}}{V_P^-(t^*)}\right)^{1 - 2\mu/\sigma^2} \Phi(d_-), \tag{5.5}
\end{equation*}
where $d_\pm = [\ln(V_{t_0}/V_P^-(t^*)) \pm \mu' T]/[\sigma\sqrt{T}]$, $T = t^* - t_0$, $t^*$ given by (5.2).

Thus, the Suicide Region is effectively closed under two conditions:
\begin{itemize}
    \item First, the initial $V_{t_0}$ must be substantially below the preemption threshold even after its descent, $V_{t_0} \ll V_P^-(t^*)$. This is a policy-relevant requirement: the mechanism's launch must precede substantial capability progress.
    \item Second, the product $\mu' T$ must be bounded above, which couples $\bar M$ to $\sigma, \mu$ and provides operational targeting. Under non-stationary volatility (the natural case if recursive self-improvement accelerates capability progress), formula (5.4) generalizes by replacing $\sigma^2 T$ with the integrated variance $\int_{t_0}^T \sigma^2(s) ds$ via the time-change representation of Brownian motion (Mörters \& Peres 2010, \S1.2). If $\sigma(t)$ grows linearly or faster, the integrated variance diverges faster than $T$, and (5.5) yields a growing probability of entry into the Suicide Region even with shrinking $T$. This is the formal residual we return to in \S7.
\end{itemize}

In other words, when $\lambda(t)$ reaches the level at which the Suicide Region is closed, alignment becomes a public good, and the strategic advantage of first deployment loses meaning. In Tan's formalism the leader's advantage is given by the parameter $S$, and at $S \to 0$ the Suicide Region is empty only when $D = 0$. But in our model, if $\pi(\tau; t^*) \approx 1$ for all relevant $\tau$, the expected damage $(1 - \pi)D \to 0$, and the effective leader--follower asymmetry vanishes. This is equivalent to simultaneous action of two interventions: de facto reduction of $D_{private}$ and endogenous emergence of the windfall effect ($S \to 1/2$) through liquidation of the leader's monopoly premium.

To operationalize this effect, it is convenient to view the regulator as a meta-participant of the race. The accumulated $R(t)$ functions as the resource base of a quasi-player with the goal of maximizing $\lambda$. This player does not compete for $V$ but competes to reach the critical level $\lambda^*$ earlier than the capability leader crosses $V_P^-(\lambda^*)$. From Lemma 1, closure occurs when $\Lambda(t) \ge L^*$, or equivalently $\lambda(t)\tau \ge L^*$ in myopic form. Accordingly
\begin{equation*}
\lambda^* = \frac{L^*}{\tau} = \frac{1}{\tau}\ln\frac{D(1-2S)}{IS} \tag{5.6}
\end{equation*}
is a constant independent of $\bar V_{\max}$. The preemption threshold at $\lambda = \lambda^*$ equals
\begin{equation*}
V_P^-(\lambda^*) = \frac{I}{(1-2S)\pi^*}, \qquad \pi^* = 1 - \frac{IS}{D(1-2S)}. \tag{5.7}
\end{equation*}

The race between the fund and the capability leader is conducted as a comparison of two times. The time from $t_0$ until $\lambda(t)$ reaches $\lambda^*$ we denote $T_\lambda$; by Theorem 1, it is bounded above by $t^* - t_0$ from (5.2) or (5.3). The time from $t_0$ until $\bar V_{\max}(t) = \mathbb{E}[\max_i V_t^i]$ reaches $V_P^-(\lambda^*)$ we denote $T_V$; in the deterministic limit $\bar V_{\max}(t) \approx V_{t_0}e^{\mu(t-t_0)}$, whence
\begin{equation*}
T_V = \frac{\Gamma}{\mu}, \qquad \Gamma = \ln\frac{V_P^-(\lambda^*)}{V_{t_0}}. \tag{5.8}
\end{equation*}

The ``fund outpaces capability leader'' condition:
\begin{equation*}
T_\lambda < T_V. \tag{P5$'$}
\end{equation*}

In myopic form with $\lambda_0 = 0$ condition (P5$'$) unfolds as
\begin{equation*}
\alpha(1-\rho)r\bar M > \frac{L^*\mu}{\tau\Gamma}. \tag{P5$'''$}
\end{equation*}

With (P5$'''$) the deterministic part is complete: within the orthogonal regime the mechanism closes the Suicide Region in finite time, and with sufficient throughput this occurs before the capability leader crosses the preemption threshold. Everything rests on (A4) --- the assumption whose explicit form has not yet been written out. The next section derives this form and proves that under its violation, scaling $\bar M$ does not save the result.

\section*{Robustness to Dual-Use and the Role of the Orthogonal Regime}
Capability and alignment are not canonically distinguishable: interpretability research is dual (Bostrom 2002; CAIF funding principles), and the scalar $\alpha$ in the pure form of \S3 is the share of the fund nominally directed into alignment but actually shifting $\mu$.

The decomposition (5.0) fixes $\rho \in [0, 1]$ as the correlation of the funded portfolio with the first principal component of ``model $\times$ benchmark'' space in the sense of Ren et al. (2024, Safetywashing). At $\rho = 0$ the portfolio is orthogonal to the capabilities component (pure differential progress; Hendrycks \& Mazeika 2022). As $\rho \to 1$ the mechanism converts cooperation surplus directly into drift and is strictly worse than inaction.

Substituting (5.0) into condition (P5$'''$) with now explicit $\mu = \mu_0 + \kappa\rho r\bar M$:
$$\alpha(1-\rho)r\bar M > \frac{L^*(\mu_0 + \kappa\rho r\bar M)}{\tau\Gamma},$$

whence after regrouping
\begin{equation*}
r\bar M\bigl[\tau\Gamma\alpha(1-\rho) - L^*\kappa\rho\bigr] > L^*_{\mu_0}. \tag{6.1}
\end{equation*}

For the existence of $\bar M > 0$ satisfying (6.1), the coefficient in brackets must be positive:
\begin{equation*}
\tau\Gamma\alpha(1-\rho) > L^*\kappa\rho \iff \rho < \rho_{crit} := \frac{\tau\Gamma\alpha}{\tau\Gamma\alpha + L^*\kappa}. \tag{6.2}
\end{equation*}

We obtain the threshold under which (A4) is operationalized. Substantively $\rho_{crit}$ depends on three dimensionless and one dimensional ratios:
\begin{itemize}
    \item The logarithmic capability gap $\Gamma$ raises $\rho_{crit}$; if deployment is far off, leakage is less dangerous.
    \item The safety gap $L^*$ lowers $\rho_{crit}$; more work for the fund, more sensitivity to leakage.
    \item The ratio $\alpha/\kappa$ is the efficiency of safety to leakage gain: raises $\rho_{crit}$ as it grows. The dimensional coefficient $\tau$ enters through the product $\tau\Gamma\alpha$.
\end{itemize}

\textit{Theorem 3 (non-rescuability by scaling).} Under (A1)--(A5) in the myopic safety approximation:
\begin{enumerate}[label=(\alph*)]
    \item If $\rho < \rho_{crit}$, there exists $\bar M_0(\rho) < \infty$ such that for all $\bar M \ge \bar M_0(\rho)$ (P5$'$) holds, with
    \begin{equation*}
    \bar M_0(\rho) = \frac{L^*_{\mu_0}}{r[\tau\Gamma\alpha(1-\rho) - L^*\kappa\rho]}. \tag{6.3}
    \end{equation*}
    \item If $\rho \ge \rho_{crit}$, for any $\bar M > 0$ (P5$'$) fails. Increasing the matchmaker's throughput does not restore the condition.
    \item $\bar M_0$ is strictly increasing in $\rho$:
    $$\frac{\partial \bar M_0}{\partial \rho} = \frac{L^*_{\mu_0}(\tau\Gamma\alpha + L^*\kappa)}{r[\tau\Gamma\alpha(1-\rho) - L^*\kappa\rho]^2} > 0.$$
\end{enumerate}
The only restoration lever is reduction of $\rho$.

\textit{Proof.} (b) At $\rho \ge \rho_{crit}$ the coefficient $\tau\Gamma\alpha(1-\rho) - L^*\kappa\rho \le 0$, the LHS of (6.1) is non-positive for any $\bar M > 0$, the RHS $L^*_{\mu_0} > 0$. Condition (P5$'$) fails. (a) At $\rho < \rho_{crit}$ the coefficient is strictly positive, (6.1) holds for $\bar M > \bar M_0(\rho)$ from (6.3). (c) Direct differentiation. $\blacksquare$

Theorem 3 is formulated in the myopic safety approximation, and its transfer to the strict hazard-rate form qualitatively changes the picture of the asymptotics of $\bar M$. In the hazard-rate form with $\lambda_0 = 0$ we have $T_\lambda \sim \sqrt{2L^*/[\alpha(1-\rho)r\bar M]} = O(\bar M^{-1/2})$ (square-root reduction due to the quadratic accumulation $\Lambda(t) = \int \lambda(s) ds$), while $T_V = \Gamma/(\mu_0 + \kappa\rho r\bar M) = O(\bar M^{-1})$ when $\kappa\rho > 0$. Then
$$T_\lambda/T_V = O(\bar M^{1/2}) \to \infty \quad \text{as } \bar M \to \infty,$$

that is, in the strict form asymptotic increase of matchmaker throughput does not satisfy (P5$'$) when $\rho > 0$ --- on the contrary, as $\bar M \to \infty$ the gap $T_V - T_\lambda$ becomes increasingly negative. Dual-use leakage in (5.0) scales linearly with $\bar M$ through the drift $\kappa\rho r\bar M$, whereas safety accumulation through the integrated $\Lambda(t)$ gives only a sublinear reduction in $T_\lambda$ (early $\lambda(s)$ are small and drag the average down), and at large $\bar M$ leakage outpaces accumulation.

Substantively the practical conclusion is preserved, and scaling $\bar M$ as a lever loses on two independent counts:
\begin{itemize}
    \item realistic limits on matchmaker throughput bound $\bar M$ above empirically
    \item in the strict hazard-rate form, asymptotic scaling is counterproductive at any $\rho > 0$ and requires pointwise calibration within a finite window.
\end{itemize}

In essence, control over $\rho$ becomes the only reliable lever, which transforms the matchmaker calibration problem into the problem of controlling $\rho$. Control is operationalized as a disbursement mandate: $M$ disburses $R(t)$ only into projects with measured $\hat\rho_{proj} \le \bar\rho$, where $\bar\rho < \rho_{crit}$ is a chosen threshold with a margin, $\hat\rho$ is the correlation of the funded work with the capabilities component according to Ren et al. (2024). This turns Bostrom's differential progress criterion from an intuitive directive into a falsifiable metric with per-grant verification. The effectiveness of the gate depends on how publicly observable $\hat\rho$ is and how verifiable disbursement decisions are, i.e., on institutional structure.

The operational gate shifts $\rho$ but does not zero it out. As of 2026 there is no canonical criterion of differential interpretability: the safety value of interpretability remains a matter of open dispute, the symmetry between detection of jailbreaks and their construction is not operationalized. Part of the funded portfolio has a lower bound $\rho_{\min} > 0$ not reducible by available methods. At first glance this is a critical vulnerability, since formulas (6.1)--(6.3) are derived for $\rho < \rho_{crit}$, and the subspace with $\rho_{\min}$ falls outside the scope of Theorems 1--3. The substantive question, however, is not ``does the subspace fall out'' but ``relative to what benchmark is the loss assessed''. The improvement bound in (6.2) is defined relative to the idealized point $\rho = 0$, and relative to it $\rho_{\min}$ does indeed bound the achievable, but in reality $\rho = 0$ is not the empirical counterfactual. As of 2026, alignment research is financed by philanthropic foundations, government grants, and lab internal budgets without unified $\hat\rho$-filtering; allocation occurs across projects with unmeasured correlation to the capabilities component, and part of this funding empirically works in the dual-use direction, which Ren et al. (2024) record as the motivation for their metric. Accordingly, the relevant benchmark for evaluating the mechanism is the status quo.

Relative to the status quo, the subspace with $\rho_{\min}$ ceases to be a subspace of pure loss. In the worst case, with a generously set $\bar\rho$ and weak measurement protocol, the mechanism reproduces the existing distribution of alignment funding: $\dot\lambda$ grows approximately as it does under current philanthropic channels. In the typical case, per-grant measurement of $\hat\rho$ provides finer resolution than the disjoint decisions of individual funds, and allocation monotonically improves relative to baseline. What we record in the formal part as a rollback to exogenous interventions ($D_{private}$, verification) on the subspace with $\rho_{\min}$ in fact means a rollback to the same instruments which on this subspace already work in the status quo by default: the guarantee of Theorems 1--3 fails, but what would be in place without it is not made worse either.

The apparent asymmetry, that undirected alignment funding is credited as a plus in the current literature, while in our analysis the same channel figures as a constraint, is an artifact of the reference point. The same flow of funds receives three different signs depending on the benchmark: relative to inaction unconditionally positive, relative to the ideal $\rho = 0$ bounded above, relative to the status quo a Pareto improvement, which turns this result from a formal one into an operational one.

\section*{Regulator Governance}
The operational gate requires institutional conditions:
\begin{itemize}
    \item $\hat\rho$ must be publicly observable
    \item disbursement decisions must be verifiable
    \item expertise to evaluate $\hat\rho$ must be embedded in decision-making; the organization's structure must be formed by its participants directly.
\end{itemize}

Transparency works as a commitment device (Schelling 1960; Kydland \& Prescott 1977): all flows $R(t)$, all allocation decisions, all participant-admission decisions are public; deviation from the declared mandate is immediately observable. This creates necessary but not sufficient conditions for resolving the Trust AI Regulation problem (Alalawi et al. 2026): transparency makes deviation observable, but coordinated response of participants (exit from the platform) itself is subject to the collective action problem, since a lab continuing to receive cooperation surplus has an individual incentive to stay even after public deviation by the regulator. Sufficiency is achieved by the same structural condition that suppresses bypass in (P4): small $N$, repeated interactions with high effective $\gamma$, and focal coordination in exclusionary behavior (Schelling 1960; Bernstein 1992; Greif 1993). Transparency of $\hat\rho$ plays a dual role here: in addition to controlling the orthogonal regime, it transforms the exit decision from a private choice into a publicly observable signal, making the coordinated response equilibrium by the same mechanism that supports (P4). Under these conditions, the de-escalating equilibrium is attainable. The bypass-suppression condition from (P4) is combined with the publicity of $\hat\rho$ in one signal: ``the mechanism works in its declared regime'' is equivalent to ``$\hat\rho \le \bar\rho$ is publicly confirmed''.

$M$'s decisions on calibration of $r^*$, allocation of $R(t)$, and admission of participants are made through a delegative-voting procedure. The literature on liquid democracy (Kahng, Mackenzie \& Procaccia 2021; Brill \& Talmon 2018; Christoff \& Grossi 2017) gives formal conditions for its stability. Participants have heterogeneous expertise in alignment relative to capabilities, the delegative structure allows peripheral participants to delegate decisions to experts, transparency of delegations preserves accountability. Among classes of governance (liquid democracy, futarchy Hanson 2013, quadratic voting Weyl 2017, sortition Landemore 2020), delegative structures form a family of focal Schelling equilibria under the conditions of transparency and voluntary participation. The same expert delegation is the institutional carrier of the $\hat\rho$ evaluation, which will be useful when adding new participants and forming the most accurate possible estimates of $\hat\rho$ through aggregation of diversified expert judgments (Hong \& Page 2004; Surowiecki 2004), which is strictly better than the operational-gate status quo relative to the disjoint decisions of individual funds. With subsequent narrowing of $\bar\rho$ as measurement standards consolidate, the distribution of funds improves monotonically relative to baseline; moreover, such a structure facilitates external and internal audits and is convenient for the application of benchmarks.

There are direct institutional analogs confirming feasibility. Self-regulatory organizations in the financial sector (DeMarzo, Fishman \& Hagerty 2005) implement the same structure of financing and legitimation without coercion. Patent pools (Lerner \& Tirole 2004) give an analog for the problem of joint IP use, with the caveat that Lerner \& Tirole enumerate direct interest of participants in blocking the rest as a condition of pool failure. In our case, this condition is mitigated by (P4) and the orthogonal regime: blocking others through $M$ requires bypassing the gate, which is observable under transparency, and exclusion from the subsequent flow of deals. A carbon tax with revenue recycling (Bovenberg \& van der Ploeg 1994) gives an analog of the double dividend: the fee simultaneously internalizes the externality and finances a public good.

Governance closes the gap between the formal mechanism and its operational implementation. Under a transparent delegative structure, the gate $\hat\rho \le \bar\rho$ works, (P4) holds, and the central results of \S5--\S6 acquire an institutional carrier. What remains to verify is robustness of the mechanism to two classes of threats unrelated to governance: nonlinear capability acceleration and transition to a full-information regime.

\section*{Non-stationary Capability and Information Regime}
Under a correctly functioning orthogonal regime and governance, two classes of residual threats require separate analysis.
\begin{itemize}
    \item Non-stationarity of capability progress: if $\mu$ and $\sigma$ grow over time, the race of two monotonically growing quantities ($\lambda$ and $\bar V_{\max}$) is decided by the ratio of their derivatives, and Theorem 1 does not automatically guarantee the fund's win.
    \item Verification systems can move the game to a full-information regime, which in the endgame may cause breakout instability.
\end{itemize}

We address both classes and show that the first partially self-corrects via Pigouvian rate recalibration, leaving only the volatility channel as an open residual, and the second adds no independent fragility on top of the first. Non-stationarity of capability decomposes into two channels:
\begin{itemize}
    \item Drift channel: when $\mu$ and $\bar V_{\max}$ grow, the Pigouvian rate is calibrated proportionally to the magnitude of the externality itself, $r^* \propto (k-1)(V_i + V_j)$ from (P3). Then $\dot\lambda = \alpha r^* M(t) \propto \bar V_{\max}(t) M(t)$, while the time to reach the preemption threshold under $\dot{\bar V}_{\max} \propto \mu\bar V_{\max}$ gives $T_V \propto \ln(V_P^*/V_{t_0})/\mu$. Both sides of (P5$'''$) scale with $\bar V_{\max}$ --- the tax base grows together with what the race is against. Pure drift acceleration does not flip the sign of (P5$'$): the drift channel self-corrects when $\rho = 0$. When $\rho > 0$ the same $\bar V_{\max}$-normalization works against the fund: leakage rides on the same base from the opposite side, and self-correction does not occur outside the orthogonal regime --- this is already formalized by Theorem 3.
    \item Volatility channel: a residual that survives \S5 and is not closed by recalibration. Under non-stationary $\sigma(t)$, crossing of $V_P^-(t)$ occurs as the first passage of a fluctuation whose intensity is uncoupled from the flow of match-deals $M(t)$ and from the accumulated $R(t)$. The breakthrough becomes a tail event not correlated with $\int M$, and monotonicity of $\lambda(t)$ does not provide protection against such a sprint; we have a bound on $\mathbb{P}[\text{first passage before } \lambda(t) \ge \lambda^*]$ under non-stationary $\sigma(t)$. This is a stochastic problem of first attainment with time-varying boundary and self-damping dynamics of $\lambda$, not covered by the deterministic race of rates in (P5$'$) and in Theorem 3. It is moved to open questions.
\end{itemize}

Here we also describe the internal tension of Pigouvian recalibration. Under continuous recalibration $r^*(\lambda(t))$ from (P3), the safety dynamics becomes a self-damping nonlinear ODE:
$$\dot\lambda(t) = \alpha M(t) D\frac{2(k-1)}{\mu}\lambda(t)e^{-\lambda(t)\tau},$$

where the factor $e^{-\lambda\tau}$ chokes growth at large $\lambda$. The linear growth from \S3 holds only under fixed $r$. This simultaneously strengthens the volatility residual (a slowing $\lambda$ wins the race against accelerating capability less well) and provides a constructive lever: fixing $r$ at the project level without downward recalibration preserves linear growth at the cost of moderate over-funding in alignment at late stages. This is an error in a benign direction. Thus, non-stationarity of capability progress narrows down to analysis of the volatility channel.

Verification systems can move the game to a full-information regime, which in the endgame accelerates the sprint to deployment --- breakout instability. A straightforward reading as ``full information requires a stricter $r^*$ and earlier bootstrap'' errs in the sign of the effect. By the transparency requirement of \S7, the state of the fund $(R(t), \lambda(t))$ is public \textit{by construction in all information regimes} --- it is not a private signal disclosed by verification, but by definition common knowledge. Rivals' capability leads $V^i$ are private without verification and public with it. The transition to full information removes the private informational advantage of rivals without changing the fund's position. The fund's relative informational position improves monotonically with growing openness.

Mechanically this sharpens the inversion of the leader's advantage described in \S5. Under incomplete information, the collapse of the preemption incentive after $\lambda(t) \ge \lambda^*$ is delayed and noised by strategic uncertainty: a player continues racing without common knowledge that rivals will not deploy unsafely. Under public $\lambda(t)$, the moment of crossing $\lambda^*$ becomes common knowledge, and the preemption game unwinds by iterated dominance immediately and in coordination. Full information accelerates breakout, but equally accelerates the coordinated stand-down on which \S5 is built; accordingly, the information regime adds no independent fragility.

After the two corrections, the residual reduces to the conjunction of the timing residual of the volatility channel with the interpretability residual of \S6: common knowledge of $\lambda(t)$ is useful exactly to the extent that it is common knowledge that $R(t)$ buys safety, not capability --- i.e., to the extent that $\hat\rho$ is publicly credible.

\section*{Open Problems}
Government programs (Manhattan-style national AI initiatives of the US and China) can in principle bypass a voluntary regulator. This is an analog of the problem of international climate negotiations (Barrett 2003) and resistance of sovereign actors to soft-law regimes. However, there is a substantial mitigation specific to the AI domain: even in the absence of $M$, a direct match between major government programs is highly unlikely under current political conditions. Export restrictions, compute controls, outbound investment controls (Executive Order 14105 and analogs), CFIUS reviews, Chinese counter-symmetric restrictions turn any R\&D cooperation into an act requiring multi-level approval with a predictably negative outcome. The public rhetoric of both sides frames AI development as geopolitical competition with a zero-sum component. State actors are de facto already in a non-cooperation equilibrium relative to each other, and the potential synergy multiplier for them is already lost --- but not converted into alignment. The threat of bypass by state actors is less than it appears on a naive reading: they do not bypass $M$ while keeping the cooperation gain, they are already outside of cooperation regardless of $M$. A more realistic picture is to view $M$ as a network superstructure over those participants for whom inter-bloc matching is possible without violating export controls. Within a single bloc (US: OpenAI, Anthropic, Google DeepMind, xAI, Meta AI), state pressure or direct financing may induce ignoring $M$, especially under liability shield in exchange for coordination in a federal structure. On this reading, this direction remains open. One should also be wary that actors with vast unilateral resources may independently accelerate capability drift, potentially exceeding the preemption threshold.

Calibration of the Pigouvian rate requires estimation of $\lambda e^{-\lambda\tau}/\mu$ and $D$, for which there is no canonical metric. This is an analog of the problem of estimating the social cost of carbon (Nordhaus 2017; Stern 2007). Robust calibration is possible through a principles-based approach with public revision. Calibration of $r^*$ is coupled with the estimate of $\rho_{crit}$ from \S6: the orthogonal-regime threshold $\bar\rho$ must be revised together with $r^*$. The operational formula from (6.3) gives the minimum matchmaker throughput under given parameters:
\begin{equation*}
\bar M_0(\rho) = \frac{L^*_{\mu_0}}{r[\tau\Gamma\alpha(1-\rho) - L^*\kappa\rho]}, \quad \rho < \rho_{crit}, \tag{9.1}
\end{equation*}
or equivalently normalized by baseline drift:
\begin{equation*}
\frac{r\bar M_0}{\mu_0} = \frac{L^*}{\tau\Gamma\alpha(1-\rho) - L^*\kappa\rho}. \tag{9.2}
\end{equation*}

For empirical calibration the following are required: $L^* = \ln[D(1-2S)/(IS)]$ from scenario estimates of $D, I, S$; $\Gamma = \ln(V_P^*/V_{t_0})$ from current market capitalization of frontier labs and (5.7); $\tau$ from researcher surveys; $\alpha$ from the historical effect of alignment investments; $\kappa$ from the average capability-spillover per dollar of interpretability research from Ren et al. (2024); $\mu_0$ from compute scaling laws; $\rho$ from the distribution of $\hat\rho$ across the portfolio, auditable per-grant. In the endgame regime as $\lambda(t) \to \lambda^*$, the Pigouvian rate takes the form $r^*_{end} = \lambda^* IS \cdot 2(k-1)/[(1-2S)\mu]$ via the substitution $e^{-L^*} = IS/[D(1-2S)]$. $D$ drops out via this relation, and $r^*_{end}$ depends only on $\lambda^*, I, S, k, \mu$. This provides a calibration point; intermediate values are interpolated via (P3) with time-varying $\lambda(t)$. The minimum subsidy $R(0)$ ensuring transition to a regime protected against bypass is the substitution $\bar M = \bar M_0(\rho)$ into computations of the expected volume of match-deals to reach the exclusion threshold (P4) as a subset of the calibration of $\bar M_0$.

Theorems 1--3 are formulated in the static formulation of thresholds, whereas in the full option game with time-varying $\lambda(t)$ the thresholds are given by an HJB problem with non-constant coefficients, and although the heuristic gives a plausible sign, a rigorous generalization of value-function convexity (Grenadier 2002) to the non-stationary strategic case is not obtained in the present work.

Also, the two irreducible robustness residuals compound, and the orthogonal regime (formula 6.2) is the only construction in the work that breaks this feedback, and its effectiveness is bounded above precisely by $\rho_{\min}$. The early-stage binding constraint, separated from rate calibration, is consolidated here: at early stages, what binds is not the size of $R(0)$ but the finding of a seed-funded portfolio within the orthogonal regime, $\hat\rho < \rho_{crit}$. A well-capitalized regulator outside the regime is strictly worse than inaction (Theorem 3); a poorly capitalized one within it remains net-positive. The philanthropic seed is optimized by $\hat\rho$, not just by volume.

\section*{Conclusion}
We demonstrate that the proposed mechanism is capable of effectively de-escalating the AGI race, operating at several levels: from the individual benefit of particular labs to global change in safety dynamics.

At the micro level, we have shown that under characteristic capitalization parameters, participation in the matchmaker's work strategically dominates direct cooperation and its absence. This is ensured within the assumptions that the costs of establishing, agreeing IP-sharing, and bilateral verification of a direct deal are high under current conditions, and the benefit of joint use of compute and data substantially outweighs the levied fee. So that the mechanism does not undermine public safety, we computed the optimal size of the Pigouvian fee. It is calibrated so that the sum of collected funds exactly covers the additional risk of catastrophe arising from acceleration of progress under cooperation. Moreover, in the conditions of repeated interaction, labs will not attempt to bypass the matchmaker, since the risk of exclusion from future profitable deals makes honest play strategically dominant.

At the macro level, we proved closure of the Suicide Region in the static formulation of thresholds in finite time and bounded above the probability of premature entry into the region before its closure. To minimize the probability of a technological sprint, the mechanism's launch must occur before progress in capabilities becomes too rapid.

\section*{References}
\begingroup
\sloppy
\begin{enumerate}[itemsep=2pt,parsep=0pt,leftmargin=15pt]
    \item Alalawi, F., Han, T. A., Zisis, I., Lenaerts, T. \& Santos, F. C. (2026). Trust AI Regulation? Discerning users are vital to build trust and effective AI regulation. \textit{Applied Mathematics and Computation}.
    \item Armstrong, S., Bostrom, N. \& Shulman, C. (2016). Racing to the precipice: A model of artificial intelligence development. \textit{AI \& Society}, \textit{31}(2), 201--206.
    \item Barrett, S. (2003). \textit{Environment and Statecraft: The Strategy of Environmental Treaty-Making}. Oxford University Press.
    \item Bernstein, L. (1992). Opting out of the legal system: Extralegal contractual relations in the diamond industry. \textit{Journal of Legal Studies}, \textit{11}(1), 115--157.
    \item Borodin, A. N. \& Salminen, P. (2002). \textit{Handbook of Brownian Motion: Facts and Formulae}. Birkhäuser.
    \item Bostrom, N. (2002). Existential risks. \textit{Journal of Evolution and Technology}, \textit{9}.
    \item Bovenberg, A. L. \& van der Ploeg, F. (1994). Environmental policy, public finance and the labour market in a second-best world. \textit{Journal of Public Economics}, \textit{55}(3), 349--390.
    \item Brill, M. \& Talmon, N. (2018). Pairwise liquid democracy. \textit{IJCAI}, 137--143.
    \item Caillaud, B. \& Jullien, B. (2003). Chicken \& egg: Competition among intermediation service providers. \textit{RAND Journal of Economics}, \textit{34}(2), 309--328.
    \item Christoff, Z. \& Grossi, D. (2017). Binary voting with delegable proxy: An analysis of liquid democracy. \textit{TARK}.
    \item Cimpeanu, T., Santos, F. C., Pereira, L. M., Lenaerts, T. \& Han, T. A. (2022). Artificial intelligence development races in heterogeneous settings. \textit{Scientific Reports}, \textit{12}, 1723.
    \item Clifton, J. \& Martin, S. Differential Progress in Cooperative AI: Motivation and Measurement. Cooperative AI Foundation seminar / working note.
    \item DeMarzo, P. M., Fishman, M. J. \& Hagerty, K. M. (2005). Self-regulation and government oversight. \textit{Review of Economic Studies}, \textit{72}(3), 687--706.
    \item Dixit, A. K. \& Pindyck, R. S. (1994). \textit{Investment under Uncertainty}. Princeton University Press.
    \item Greif, A. (1993). Contract enforceability and economic institutions in early trade. \textit{American Economic Review}, \textit{83}(5), 524--548.
    \item Grenadier, S. R. (2002). Option exercise games: An application to the equilibrium investment strategies of firms. \textit{Review of Financial Studies}, \textit{15}(3), 691--721.
    \item Han, T. A., Pereira, L. M., Santos, F. C. \& Lenaerts, T. (2020). To regulate or not: A social dynamics analysis of an idealised AI race. \textit{JAIR}, \textit{69}, 881--921.
    \item Han, T. A., Lenaerts, T., Santos, F. C. \& Pereira, L. M. (2021). Voluntary safety commitments provide an escape from over-regulation in AI development. \textit{Technology in Society}. arXiv:2104.03741.
    \item Hanson, R. (2013). Shall we vote on values, but bet on beliefs? \textit{Journal of Political Philosophy}, \textit{21}(2), 151--178.
    \item Hendrycks, D. \& Mazeika, M. (2022). X-Risk Analysis for AI Research / Pragmatic AI Safety. arXiv:2206.05862.
    \item Hong, L. \& Page, S. E. (2004). Groups of diverse problem solvers can outperform groups of high-ability problem solvers. \textit{PNAS}, \textit{101}(46), 16385--16389.
    \item Kahng, A., Mackenzie, S. \& Procaccia, A. D. (2021). Liquid democracy: An algorithmic perspective. \textit{JAIR}, \textit{70}, 1223--1252.
    \item Karatzas, I. \& Shreve, S. E. (1991). \textit{Brownian Motion and Stochastic Calculus}. Springer.
    \item Kydland, F. E. \& Prescott, E. C. (1977). Rules rather than discretion. \textit{Journal of Political Economy}, \textit{85}(3), 473--491.
    \item Landemore, H. (2020). \textit{Open Democracy}. Princeton University Press.
    \item Lerner, J. \& Tirole, J. (2004). Efficient patent pools. \textit{American Economic Review}, \textit{94}(3), 691--711.
    \item Mörters, P. \& Peres, Y. (2010). \textit{Brownian Motion}. Cambridge University Press.
    \item Nordhaus, W. D. (2017). Revisiting the social cost of carbon. \textit{PNAS}, \textit{114}(15), 1518--1523.
    \item O'Keefe, C., Cihon, P., Garfinkel, B., Flynn, C., Leung, J. \& Dafoe, A. (2020). The windfall clause: Distributing the benefits of AI for the common good. \textit{Centre for the Governance of AI}.
    \item Pigou, A. C. (1920). \textit{The Economics of Welfare}. Macmillan.
    \item Ren, R., Basart, S., Khoja, A., Gatti, A., Phan, L., Yin, X., Mazeika, M., Pan, A., Mukobi, G., Kim, R. H., Fitz, S. \& Hendrycks, D. (2024). Safetywashing: Do AI Safety Benchmarks Actually Measure Safety Progress? \textit{NeurIPS 2024 Datasets \& Benchmarks Track}. arXiv:2407.21792.
    \item Rochet, J.-C. \& Tirole, J. (2003). Platform competition in two-sided markets. \textit{JEEA}, \textit{1}(4), 990--1029.
    \item Sandmo, A. (1975). Optimal taxation in the presence of externalities. \textit{Swedish Journal of Economics}, \textit{77}(1), 86--98.
    \item Schelling, T. C. (1960). \textit{The Strategy of Conflict}. Harvard University Press.
    \item Stern, N. (2007). \textit{The Economics of Climate Change: The Stern Review}. Cambridge University Press.
    \item Surowiecki, J. (2004). \textit{The Wisdom of Crowds}. Doubleday.
    \item Tan, D. (2025). The Suicide Region: Option Games and the Race to Artificial General Intelligence. Working paper.
    \item Weeds, H. (2002). Strategic delay in a real options model of R\&D competition. \textit{Review of Economic Studies}, \textit{69}(3), 729--747.
    \item Weyl, E. G. (2017). The robustness of quadratic voting. \textit{Public Choice}, \textit{172}(1--2), 75--107.
    \item Williamson, O. E. (1979). Transaction-cost economics: The governance of contractual relations. \textit{Journal of Law and Economics}, \textit{22}(2), 233--261.
    \item Enabling Frontier Lab Coordination to Mitigate AI Safety Risks (2025). arXiv:2511.08631.
\end{enumerate}
\endgroup

\end{document}